\documentclass[prl,a4paper,twocolumn,showpacs,superscriptaddress]{revtex4}
\usepackage{graphicx}
\usepackage{amsmath,amssymb,braket}
\usepackage{hyperref}
\usepackage{color}

\newcommand{\hidden}[1]{}

\begin{document}
\preprint{xyz}

\title{Spin-orbit fields in asymmetric (001) quantum wells}
\author{P.~S.~Eldridge}
\email[Correspondence to: ]{eldridge@nano.uni-hannover.de} %
\author{J.~H{\"u}bner}
\author{S.~Oertel}
\affiliation{Institute for Solid State Physics, Leibniz University
of Hannover, Appelstr. 2, 30167 Hannover, Germany}
\author{R.~T.~Harley}%
\affiliation{School of Physics and Astronomy, University of
Southampton, Southampton, SO17 1BJ, UK}
\author{M.~Henini}%
\affiliation{School of Physics and Astronomy, University of
Nottingham, Nottingham, NG7 4RD, UK}
\author{M.~Oestreich}%
\affiliation{Institute for Solid State Physics, Leibniz University
of Hannover, Appelstr. 2, 30167 Hannover, Germany}%

\date{\today}

\begin{abstract}

We measure simultaneously the in-plane electron g-factor and spin
relaxation rate in a series of undoped inversion-asymmetric (001)-oriented
GaAs/AlGaAs quantum wells by spin-quantum beat spectroscopy. In
combination the two quantities reveal the
absolute values of both the Rashba and the Dresselhaus
coefficients and prove that the Rashba coefficient can be
negligibly small despite huge conduction band potential gradients which
break the inversion symmetry. The negligible Rashba coefficient is a
consequence of the 'isomorphism' of conduction and valence band
potentials in quantum systems where the asymmetry is solely
produced by alloy variations.

\end{abstract}


\maketitle

Symmetry is a thread which runs through all of physics and symmetry
reduction discloses basic physical
principles. In this letter, we employ crystallographically engineered
symmetry reduction to study the intricate effects of spin-orbit
interaction on the electron spin in semiconductor nanostructures.
Symmetry reduction is an especially powerful tool in
semiconductor physics because the variety of crystallographic
directions combined with 
bandgap engineering allow enormous freedom.

The interplay between structure, symmetry and electron spin in semiconductors directly affects
the spin relaxation rate $\Gamma_s$ and the effective electron
Land{\'e}-factor $g$. Early studies of $\Gamma_s$ and
$g$ were focused on bulk zincblende material where both
entities are isotropic \cite{OpticalOrientation}. Subsequently,
the reduction in symmetry from T$_d$ to D$_{2d}$ symmetry in
symmetrical (001)-oriented quantum wells (QWs) was shown to give rise to
anisotropy between the in-plane (x,y) and the out of plane (z)
directions \cite{Ivchenko1992,Dyakonov1986}. Further reduction in
symmetry to C$_{2v}$ is achieved in (001) quantum wells by
removing the mirror symmetry of the quantum well potential and
allows an in-plane, two-fold symmetric anisotropy of
both $\Gamma_s$ \cite{Averkiev1999} and  $g$
\cite{Kalevich1993}.

Fundamentally, $\Gamma_s$ and $g$ are both determined by spin-orbit interaction but the basic mechanisms for their
anisotropies are quite different. Theoretically the
in-plane anisotropy of $g$ is proportional to the asymmetry
of the electron wavefunction in the growth direction with proportionality constant given by the Dresselhaus or bulk inversion asymmetry (BIA)
spin-splitting coefficient $\gamma$ \cite{Dresselhaus1955,
Kalevich1993}. In contrast, $\Gamma_s$ is in many cases dominated
by the Dyakonov-Perel spin-relaxation mechanism and the related
in-plane anisotropy depends on the ratio
$(\alpha/\beta)$ of the Rashba structural inversion asymmetry
(SIA) to the BIA spin
splitting \cite{Averkiev1999}. The SIA component is determined in a rather subtle way by the asymmetry of the \emph{structure} along the growth direction \cite{Winkler2003, Eldridge2010_PRB82}. 

In this letter, we determine the absolute value of both the Rashba and Dresselhaus coefficients for a series of quantum well structures by simultaneously measuring the in-plane anisotropy of $\Gamma_s$ and $g$ by spin
quantum beat spectroscopy \cite{Heberle1994}. The specially designed undoped (001) quantum well samples, with reduced C$_{2v}$ symmetry but {\em without} external electric
fields, illustrate clearly the different origins of the two anisotropies as they possess a strong anisotropy of $g$ and nearly negligible anisotropy of $\Gamma_s$.

Anisotropies of $\Gamma_s$ and $g$ have been measured
previously in symmetrically grown quantum wells in an external
electric field \cite{Oestreich1996a, Larionov2008} but the
decisive simultaneous evaluation of Dresselhaus and Rashba
components has not been carried out so far. Hanle experiments in undoped asymmetric quantum wells without an applied electric field have revealed a strong in-plane anisotropy of the Hanle depolarization
curve \cite{Averkiev2006} but such measurements are unable to distinguish between
anisotropy of $\Gamma_s$ and $g$ \cite{Hanle1924}. Recently,
Ganichev and co-workers introduced a seminal technique that
uses the angular distribution of the spin-galvanic effect and
therewith measured the {\em ratio} of the Rashba and Dresselhaus
coefficients in doped quantum wells\cite{Ganichev2002,
Ganichev2004}. Salis and co-workers developed a technique that in principle yields the absolute values of the coefficients in doped structures by optically monitoring the angular dependence
of the electrons' spin precession \cite{Meier2007}. However as the calculation of electric fields in these samples is complicated the values of the coefficients can be overestimated \cite{Studer2009}. 

We first summarise the theoretical mechanisms for $g$ and $\Gamma_s$ anisotropy \cite{Averkiev1999, Kalevich1993}.  For $g$ a small magnetic field in $x$-direction $B_x$
deflects the rapid zero-point motion of an electron quantized in
$z$-direction and yields a change of momentum in y-direction. This
additional momentum $\delta p_{y}$ changes the effective Rashba
$\mathbf{ \Omega_{R}}$ and Dresselhaus $\mathbf{ \Omega_{D}}$
precession vectors which read for (001) quantum wells in zinc-blend
crystals
\begin{equation}\label{Eq:OmegaDR}
  \mathbf{\Omega} _R (\mathbf{p}) = \alpha/\hbar^2 \left(
{\begin{array}{*{20}c}
   {p_y }  \\
   { - p_x }  \\
   0  \\
\end{array}} \right)\quad
\mathbf{\Omega} _D(\mathbf{p})  = \beta/\hbar^2 \left(
{\begin{array}{*{20}c}
   {-p_x }  \\
   {p_y }  \\
   0  \\
\end{array}} \right)
\end{equation}
where $\alpha$ and $\beta$ are coefficients and $p_{x,y,z}$ are the components of the electron momentum.
Inspection of Eq. \ref{Eq:OmegaDR} shows that the Rashba term converts $\delta p_{y}$ into an
additional magnetic field which is parallel to the external
magnetic field $B_x$ and thereby alters the diagonal component of the g-tensor ($g_{xx}=g_{yy}$). By contrast, the Dresselhaus term $\mathbf{ \Omega_{D}}$
converts $\delta p_{y}$ to an additional magnetic field in y-direction, i.e. perpendicular to $B_x$ and thereby generates an
 off-diagonal component $g_{xy}$. A rigorous theoretical treatment yields \cite{Kalevich1993}
\begin{equation}\label{eq:gxy}
g_{xy}=g_{yx}=(2\gamma e/\hbar^3\mu_B)\left(\left\langle
p^2_z\right\rangle\left\langle z\right\rangle-\left\langle p^2_z
z\right\rangle\right),
\end{equation}
where $\mu_B$ is the Bohr magneton and $\left\langle\right\rangle$ represents an expectation value for the electron wavefunction. The
two terms in Eq. \ref{eq:gxy} cancel and $g_{xy}$ vanishes if the electron wavefunction is symmetric. The anisotropy of the g-tensor is thus proportional to the Dresselhaus coefficient $\gamma$ and determined by 
asymmetry of the electron wavefunction which may be induced by asymmetry of the confining (conduction band)
potential for the electrons. The effective g-factor for magnetic field oriented at angle $\phi$ to the (110) axis in the quantum well plane is given by 

\begin{equation}
 g(\phi)=-\sqrt{g_{xx}^{2}+g_{xy}^{2}+2g_{xx}g_{xy}\sin(2\phi)}.
 \label{equation_g}
\end{equation}

For the spin relaxation which is dominated by the Dyakonov-Perel mechanism the rate in the quantum well plane $\Gamma_{s}^{xy}(\phi)$ is proportional to $\left\langle\mathbf{\Omega^2}\right\rangle$ where $\mathbf{\Omega}=\mathbf{\Omega_D} + \mathbf{\Omega_R}$, the sum of Dresselhaus and Rashba components. It will be anisotropic as a result of interference of the components and is given by \cite{Averkiev1999}

\begin{equation}
   \Gamma_{s}^{xy}(\phi)=\frac{C}{2}(\alpha^2+\beta^2+2\alpha\beta\sin(2\phi)),\nonumber\\
\end{equation}
where $C$ is a constant which depends on the in-plane electron momentum relaxation time which is not well known in general. Thus, the spin relaxation rate anisotropy gives the ratio $\alpha/\beta$, where $\beta=<p_z^2>\gamma/\hbar^2$.

Experimentally, we measure the electron spin relaxation rate along the growth direction (z) for a magnetic field applied in the quantum well plane. The magnetic field causes rapid Larmor precession of the electron spins about the magnetic field and the measured relaxation rate is given by the average of $\Gamma_{s}^{z} =C (\alpha^{2}+\beta^{2})$ and $\Gamma_{s}^{xy} (\phi)$ \cite{Larionov2008}:
\begin{eqnarray}\label{eq:TauS}
    \Gamma_s(\phi)=\frac{1}{2}(\Gamma_s^z+\Gamma_s^{xy}(\phi))=
    D\left[1+\left(\frac{\alpha}{\beta}\right)^{2}+\frac{2\alpha}{3\beta}\sin(2\phi)\right]
\end{eqnarray}
where $D=3C\beta^2/4$. Therefore measurement of both anisotropies yields simultaneously the absolute values of $\alpha$ and $\beta$. It is interesting to note that spin relaxation rate anisotropy has the same form as the g-factor but with $\beta$ replacing $g_{xx}$ and $\alpha$ replacing $g_{xy}$.

\begin{figure}[tbp]
  \centering
  \includegraphics[width=7.5cm]{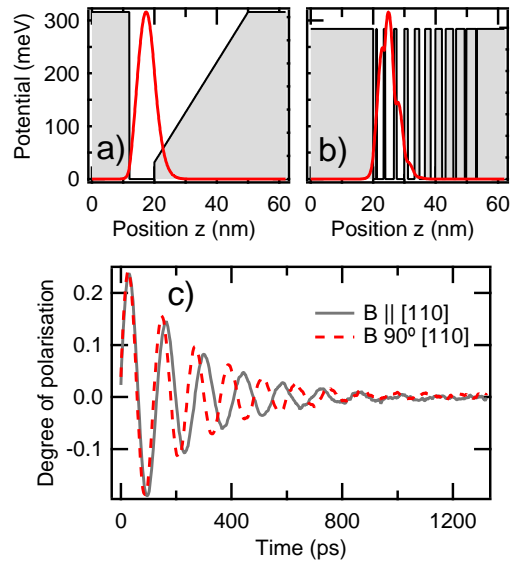}
\caption{(Color online) Conduction band potential profile and 
numerical calculated electron wavefunction for the $n=1$ states
for a) sample A and b) sample B. c) The measured spin quantum beats at 125\,K for sample A for 3\,T in-plane magnetic field clearly showing the
different electron g-factors for B$\|[110]$ and $[1\bar 10]$ and similar spin relaxation times.} \label{Fig1}
\end{figure}

The samples are four molecular beam epitaxy grown, (001)-oriented GaAs/AlGaAs multiple quantum
wells with varying asymmetry. Sample A comprises 5 repeats of a
12\,nm Al$_{0.4}$Ga$_{0.6}$As barrier, an 8\,nm GaAs quantum well
followed by a 30\,nm alloy layer where the aluminium concentration
is varied linearly from 0.04 to 0.4. Samples B-D are equivalent
structures but the one sided potential gradient is in the quantum well and 
has been grown as digital alloy with conduction band gradients
equivalent to an electric field of 100\,kV/cm for sample B,
50\,kV/cm for sample C, and 25\,kV/cm for sample D. Figure 1 shows
the calculated n=1 electron states for samples A and B obtained by
numerical solution of the Schr{\"o}dinger equation. The calculated
confinement energies for electrons in samples A to D are 34\,meV,
91\,meV, 61\,meV and 37\,meV, respectively.

The samples are mounted on a rotation stage in a liquid helium
flow cryostat in a superconducting magnet with the magnetic field
oriented in Voigt geometry. The rotation axis corresponds to the
growth axis of the sample and is parallel to the direction of
excitation. Spin oriented electrons are optically created by
circularly polarized picosecond pulses from a mode-locked
Ti:Sapphire laser with a repetition rate of 80\,MHz, a laser
wavelength of 740\,nm and a pulse intensity yielding
excitation density $\approx 2\times10^{10}$\,cm$^{-2}$. After excitation
the carrier momentum distribution rapidly thermalizes by emission
of phonons and scattering with other carriers and the holes lose their spin orientation
within the momentum relaxation time due to
strong valence band mixing and $k$ dependent spin splitting. The
polarized luminescence is spectrally and temporally resolved by a
spectrometer and a synchroscan streak camera with two-dimensional
readout which provides a resolution of 0.5\,nm and 8\,ps,
respectively. The degree of circular polarizations of the PL,
which is proportional to
the electron spin polarization, is measured by a switchable liquid
crystal retarder and a polarizer.

\begin{figure}[tbp]
  \centering
  \includegraphics[width=7.5cm]{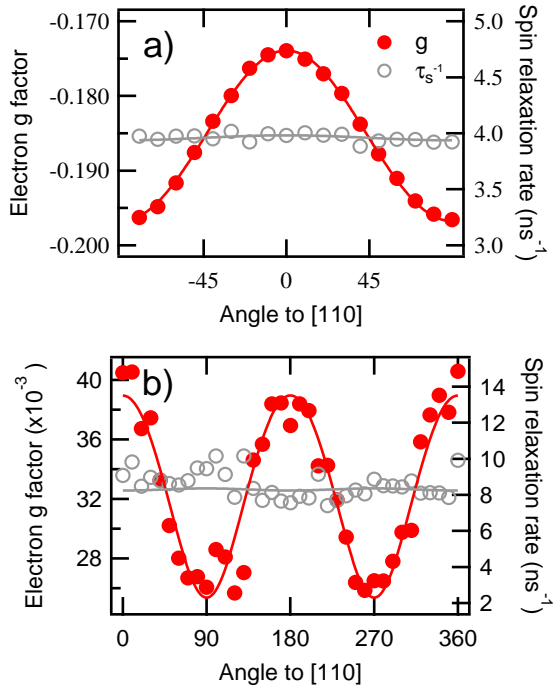}
\caption{(Color online) Extracted spin relaxation rate and electron g-factor for different
magnetic field orientations from fits to the spin quantum beat
measurements for a) sample A at 125\,K and b) sample B at 25\,K. 
}\label{Fig:AngleDep}
\end{figure}

Figure~\ref{Fig1}(c) depicts the time evolution of the
degree of circular polarization for sample~A at 3\,T and 125\,K
for an in-plane magnetic field $B$ along [110] and [1$\bar{1}$0]
directions. The observed oscillations are electron spin quantum
beats the frequency being
$\omega_L=g\mu_{B}\hbar^{-1}B$ and so a direct measure of
$g$ for the particular magnetic field direction \cite{Heberle1994}. Measurements of beats in $<S_z>$ in this way do not yield
the sign of $g$ but a comparison with previous measurements
on symmetric QWs identifies that $g$ is negative for samples
A, C, and D and positive for sample B \cite{Snelling1991, Oestreich1996a}. The two
clearly distinct oscillation frequencies in
Fig.~\ref{Fig1}(c) directly demonstrate the in-plane $g$
anisotropy whereas the nearly identical decay of the two polarisation transients indicate that $\Gamma_s$ is very nearly isotropic.

\begin{figure}[tbp]
  \centering
  \includegraphics[width=7.5cm]{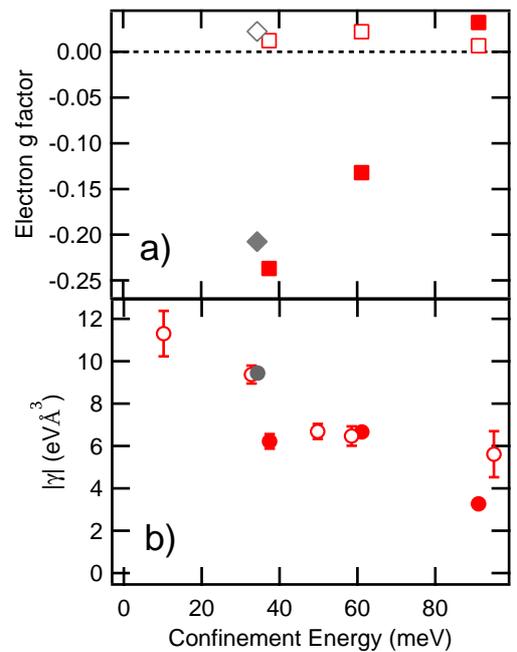}
  \caption{a) (Color online) Variation in $g_{xx}$ (solid squares) and $g_{xy}$
  (open squares) with confinement energy for samples B-C (5-25\,K) and
  sample A (25\,K) (grey diamonds). b) Experimental values of the
  Dresselhaus spin splitting constant against
  confinement energy (solid circles); open circles correspond to
  data from Ref.~\cite{Leyland2007_76}}\label{fig:gxxgyy_gamma}
\end{figure}

Figure~\ref{Fig:AngleDep} shows in more detail the dependence
of $g$ and $\Gamma_s$ on the direction of magnetic field in samples A and B.
The black (red online) solid curves in Fig.~\ref{Fig:AngleDep} depict fits of the anisotropy of
$g$ using Eq. \ref{equation_g} which directly yield both $g_{xx}$ and
$g_{xy}$. The diagonal components of the $g$-tensor
$g_{xx}=g_{yy}$ have been previously investigated
in symmetrical quantum wells where the dependence on well width,
i.e., confinement energy and barrier penetration, is well
described by k$\cdot$p theory \cite{Snelling1991,Hannak1995}. The
solid squares in figure~\ref{fig:gxxgyy_gamma}(a) show
$g_{xx}$ for all four samples confirming a similar strong dependence of $g_{xx}$ on
confinement energy for asymmetric QWs. The open squares in
Fig.~\ref{fig:gxxgyy_gamma}(a) show $g_{xy}$ and these values yield by Eq.~\ref{eq:gxy}
the dependence of the Dresselhaus spin splitting constant $\gamma$ on
confinement energy (solid dots in Fig.~\ref{fig:gxxgyy_gamma}(b)).
The excellent agreement with data from Ref.~\cite{Leyland2007_76} illustrates clearly 
that $g_{xy}$ provides an accurate measure of $\gamma$ in asymmetric (001) quantum wells. The remaining deviations
of $\gamma$ from the trend probably result from differences between the
actual and the nominal sample structures which lead to uncertainties in the calculation of the wavefunction
asymmetry. The distinct decrease of $\gamma$ with confinement
energy is expected from k$\cdot$p theory and has similar origin
to the change of $g_{xx}$ with confinement energy in
Fig.~\ref{fig:gxxgyy_gamma}(a) \cite{Leyland2007_76}. 

\begin{figure}[tbp]
  \centering
  \includegraphics[width=\columnwidth]{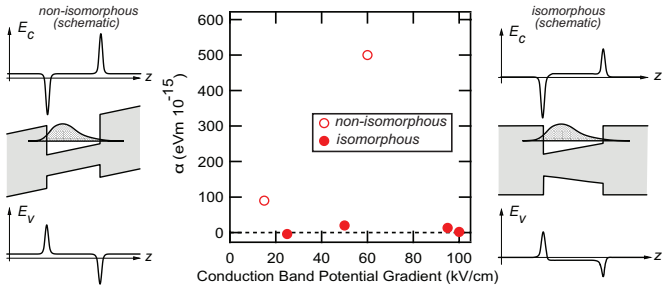}
  \caption{(Color online) Extracted values of the Rashba spin-orbit constant vs. conduction band potential gradient for samples A-D (solid circles). The open circles show values for a built-in Hartree electric field \cite{StichPRB2007} of $\sim$15 kV/cm in an n-modulation doped structure and for
an externally applied electric field of 60\,kV/cm in an undoped (110)-oriented MQW sample \cite{Eldridge2008}. Right and left panels show schematic potential profiles and electron probability density (middle) and effective electric field for conduction (top) and valence bands (bottom) (after \cite{Winkler2003}). For bound electrons, Ehrenfest's theorem forces the expectation value of the effective field in the conduction band to vanish. For an 'isomorphous' structure (right panel) the expectation value in the valence band will also vanish but \emph{not} for a 'non-isomorphous' structure (left panel), giving zero (finite) SIA spin splitting for the former (latter) \cite{Eldridge2010_PRB82}.}\label{fig:alpha}
\end{figure}

Next, we study in detail the anisotropy of the spin relaxation rate. The open circles in
Fig.~\ref{Fig:AngleDep}(a) and (b) depict $\Gamma_s(\phi)$ for sample A and B respectively and the grey
solid curves are fits according to Eq.~\ref{eq:TauS}. Additional temperature and density dependent measurements confirm
that the Dyakonov-Perel spin relaxation mechanism dominates $\Gamma_s$. The measurements clearly show that there is almost no in-plane anisotropy of $\Gamma_s$ and therefore $\alpha$ is close to zero even though the potential gradients in both samples are large ($>90$kV/cm).

Figure \ref{fig:alpha} compares $\alpha$ in our samples (solid circles) with previous experiments
in external and internal (Hartree) electric fields (open circles) \cite{Eldridge2008, StichPRB2007}. The
comparison of the measurements clearly show that the Rashba spin
splitting in AlGaAs heterostructures is large even for a modest external (or internal) electric field but negligibly small in the case of asymmetries produced by alloy variation. Although allowed to be non-zero by the $C_{2v}$ symmetry of the samples, the values of $\alpha$ which are required to fit the present data are zero within experimental uncertainties; they show both positive and negative values with no clear trend as a function of potential gradient and the fitted value of $\alpha/\beta$ is in all cases less than 0.1. The measurements push down by an order of magnitude the previous upper limit of Rashba spin splitting observed in samples with asymmetry from alloy variation \cite{Eldridge2010_PRB82}. The small values of $\alpha$ are a direct consequence of the 'isomorphous' band edges, that is the conduction and valence band potentials are related by a constant factor. This is due to the fact, that  the expectation value of the effective electric field always vanishes in the conduction band due to Ehrenfest's theorem \cite{Winkler2003} \emph{and} in 'isomorphous' structures as illustrated in the right hand panel of figure 4, will also vanish in the valence band and it is the latter which determines the spin splitting.

In conclusion, we have determined simultaneously the absolute
values for the Dresselhaus and the Rashba spin-orbit interaction
in undoped low-symmetry (001) quantum wells. All samples show a
distinctive anisotropy of the electron g-factor but essentially
isotropic spin relaxation rates. This difference highlights the
different origins of the two phenomena; the first is a measure of
the conduction electron wavefunction asymmetry and the latter a measure of the
expectation value of the valence band potential on conduction
bands states. Although, a one sided-gradient of the conduction
and/or the valence band leads in general to a finite Rashba
spin-orbit interaction, the experiment proves that isomorphism of
valence and conduction band in GaAs/AlGaAs quantum wells proscribe
a sizeable, gradient-induced Rashba spin-orbit splitting.

We thank K. K{\"o}hler for providing us with the samples
and W.~W.~R{\"u}hle for helpful discussions.  We gratefully
acknowledge financial support from Engineering and Physical Sciences Research Council (EPSRC) and from the Deutsche
Forschungsgemeinschaft in the framework of the priority programm
"SPP 1285 - Semiconductor Spintronics" and the excellence cluster
"QUEST - Center for Quantum Engineering and Space-Time Research".

\bibliographystyle{pr}
\bibliography{gfactorPapers}

\end{document}